\documentclass{article}

\usepackage{amsfonts, amsmath}
\usepackage{setspace, graphicx}
\usepackage{hyperref, epstopdf}
\usepackage{cite}

\begin{document}

\title{\textbf{Cupolets and a chaotic analog of entanglement\footnote{Paper presented at the Dynamics Days Conference, Baltimore, M.D., January 4-7, 2012}}}
\author{Matthew A. Morena and Kevin M. Short\thanks{Corresponding author. Email: \href{Kevin.Short@unh.edu}{\nolinkurl{Kevin.Short@unh.edu}}} \\
\normalsize{\emph{Integrated Applied Mathematics Program, University of New Hampshire, Durham, NH 03824, USA}} \\[-3ex]
\date{\normalsize January 4, 2012}}

\maketitle

\begin{abstract}
The realization that chaotic systems can be controlled has inspired much experimental and theoretical work seeking to stabilize the unstable periodic orbits that are typically located densely on an attractor.  One particular method, adapted from a control technique of Hayes, Grebogi, and Ott, uses small perturbations and allows for the creation of thousands of periodic orbits.  These (stabilized) chaotic unstable orbits are known as \emph{cupolets} (\emph{\underline{C}haotic, \underline{U}nstable, \underline{P}eriodic, \underline{O}rbit-\underline{LETS}}).  Cupolets exhibit the interesting property that a given set of controls will uniquely identify a cupolet, independent of the initial condition.  Practical applications of cupolets already include data compression, secure communication, and image processing.  Other recent work has shown that cupolets can be combined with graph theory and Dijkstra’s shortest path algorithm to move efficiently around an attractor, by switching from one cupolet to another.  The primary focus of this paper is to describe a chaotic analog of entanglement that can be found between cupolets.  The interaction between cupolets is assumed to be managed by an exchange function, and it has been found that cupolet pairs can become entangled in the sense that they mutually stabilize each other.  Examples of entangled cupolets will be presented and the paper will conclude with some discussion of future research directions.

\end{abstract}




\section{Introduction}
\label{sec:introduction}

While chaotic systems have been both experimentally and numerically realized in various research settings for some time, it is more recently that physicists and engineers have started synthesizing systems that exploit their properties.  In particular, the dense set of unstable, periodic orbits that chaotic systems typically admit around an attract is a rich source of qualitative information about the dynamical system. Several control schemes have been designed to extract and stabilize these orbits \cite{Hayes1993, Hayes1994}.  In this paper, we discuss an adaptation of one control method for stabilizing the unstable periodic orbits of chaotic systems.  The resulting orbits are known as \emph{cupolets}, (\emph{\underline{C}haotic, \underline{U}nstable, \underline{P}eriodic, \underline{O}rbit-\underline{LETS}}), and their abundance has been utilized in a variety of theoretical and practical applications \cite{Parker1999, Short1999, Short2005a, Short2005b, Zarringhalam2007, Johnson2009}.

Meanwhile, the implications of the uncertainty principle and the linearity of the Schrodinger equation had previously confined chaotic behavior only to classical domains.  Recent observations of chaotic behavior at the quantum level, however, indicate that dynamical production of quantum entanglement is influenced by the chaotic dynamics of the underlying system \cite{Chaudhury2009, Ghose2008, Oskay2001}.  Current developments in quantum computing further support the suspicion that the effects of chaos on entanglement may be significant \cite{Benenti2001}. These investigations suggest that these two research fields may indeed overlap more than had been believed.

For example, in \cite{Chaudhury2009}, entanglement in the purely quantum sense is both numerically and experimentally observed to be a reliable signature of classical chaos.  Under a particular choice of input parameters, the dynamics of the popular ``kicked top'' model reveal a few islands of regular trajectories surrounded by a chaotic sea.  The experimental realization of this simulation is a coupling of the electron and nuclear spins of laser-cooled Caesium ($^{133}$Cs) atoms.  Linear entropy is chosen as the entanglement measure and it was observed that greater entanglement corresponds to initial states found in the chaotic sea compared with weaker entanglement found in the regular regions.  Sensitivity to perturbation was also detected in the overlap decay of two quantum states evolving from slightly different initial conditions: the decay transpired at different rates depending on whether the evolution began from a chaotic versus regular region.  These are all indicators of chaotic behavior in a quantum setting.

In this paper, we propose an analog of entanglement in chaotic systems that we have observed between cupolets.  We do not restrict our interpretation of entanglement to its quantum context.  Rather, we regard two cupolets as entangled once a two-way coupling exists between them, and the properties of cupolet stabilization imply that the entanglement will be broken if the interaction is disturbed.  This property seems to be analogous to breaking entanglement in quantum systems.

In the implementation of cupolet entanglement described in this paper, pairs of cupolets interact by way of an exchange function.  The exchange function serves as an intermediary, and it has been found that cupolet pairs can become entangled in the sense that they mutually stabilize each other.  Creating this sort of general entanglement in a classical, nonlinear setting may ultimately be physically realizable because as a chaotic system evolves it visits its entire attractor, including visiting extremely closely its unstable periodic orbits.  Hence, if the exchange function represents a physical property of the medium or environment, then the necessary conditions for entanglement could exist.  Then, if two such systems were to interact in an environment which behaves like an exchange function, the two systems could become chaotically entangled.

The paper is organized as follows.  In Section~\ref{sec:cupolets}, we briefly discuss some interesting properties of cupolets.  We illustrate this with the double scroll system, but many other chaotic systems may be used.  We outline our entanglement scheme in Section~\ref{sec:entangled_cupolets}, and provide a few examples of exchange functions.  More complete details will be included in a future paper \cite{Morena2012}.  Then, we illustrate our results with figures of entangled cupolets.  We conclude the paper by identifying the anticipated directions of our future research and by summarizing our results.



\section{Cupolets}
\label{sec:cupolets}
 
The theory behind cupolets and their applications has been documented in \cite{Zarringhalam2007, Zarringhalam2008, Johnson2009}.  In this section, we thus only briefly describe a way of generating cupolets and then discuss their practical applications. 

\subsection{Generating Cupolets}
\label{subsec:generating_cupolets}

The method is adapted from a chaotic communication scheme developed by Hayes, Grebogi, and Ott (HGO) \cite{Hayes1993, Hayes1994} which uses small controls to induce a chaotic system to produce a signal bearing desired binary, bit stream information.  Parker and Short \cite{Parker1999, Short1999} later adapted this control scheme for secure communication by adding microcontrols.  They discovered that repeatedly applying a control code causes a trajectory to close up on itself and stabilize onto a unique periodic orbit, independent of its initial condition.  The controls that we use are arbitrarily small and, as a consequence, do not significantly alter the topology of the orbits on the chaotic attractor.

\begin{figure}[!ht]
	\begin{centering}
		\includegraphics[width=380pt]{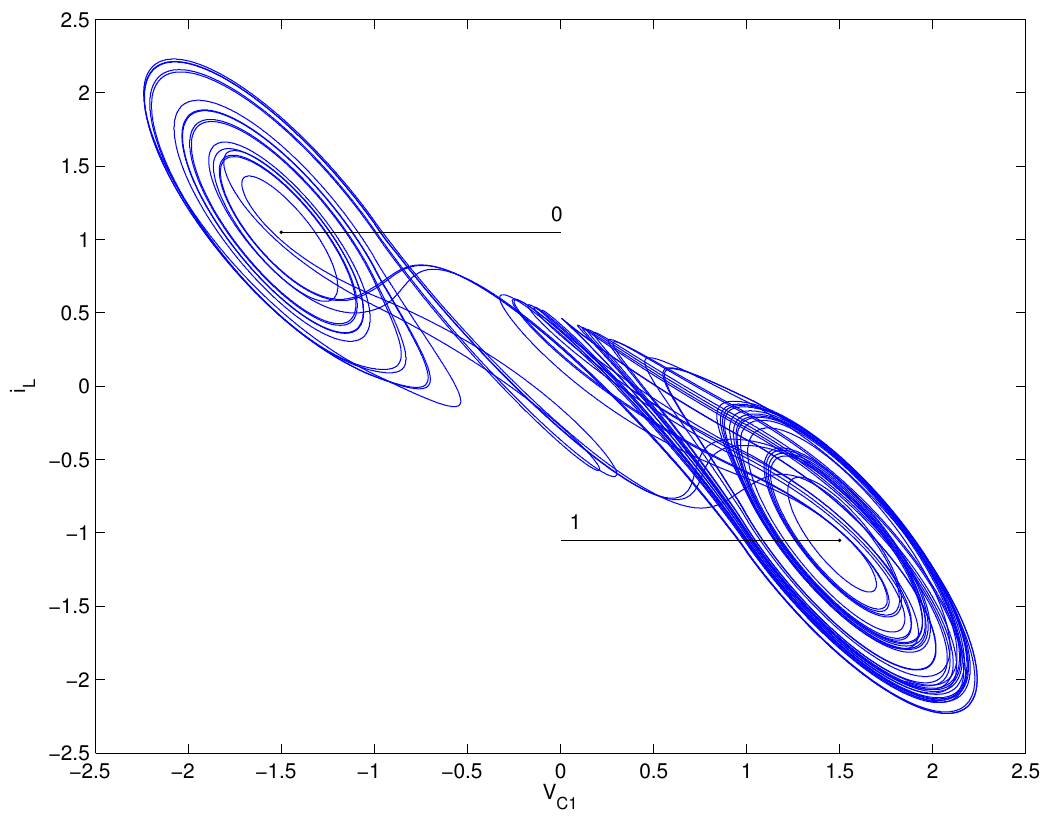}
		\caption{Double scroll oscillator showing the control surfaces.}
		\label{fig:DS_Attractor}
	\end{centering}
\end{figure}

In their paper, HGO controlled the double scroll oscillator, also known as Chua's oscillator, by using small perturbations applied on a control surface to steer trajectories around each of the two lobes of the attractor.  The differential equations describing the double scroll oscillator are:  
\begin{eqnarray*}
	C_{1}\dot{v}_{C_{1}} & = & G(v_{C_{2}}-v_{C_{1}})-g(v_{C_{1}}), \\ 
	C_{2}\dot{v}_{C_{2}} & = & G(v_{C_{1}}-v_{C_{2}})+i_{L}, \\
	L\dot{i}_{L} & = & -v_{C_{2}},
\end{eqnarray*}
where
\begin{eqnarray*}
	g(v) = \left\{ 
	\begin{array}{l}
		m_{1}v, \\ 
		m_{0}\left( v+B_{p} \right) - m_{1}B_{p}, \\ 
		m_{0}\left( v-B_{p} \right) + m_{1}B_{p},
	\end{array}
	\begin{array}{r}
		\textrm{if} \\ 
		\textrm{if} \\ 
		\textrm{if}
	\end{array}
	\begin{array}{l}
		\vert v \vert \leq B_{p}, \\
		v \leq -B_{p}, \\ 
		v \geq B_{p},
	\end{array}
	\right. 
\end{eqnarray*}
with parameter values $C_{1}=\frac{1}{9}$, $C_{2}=1$, $L=\frac{1}{7}$, $G=0.7$, $m_{0}=-0.5$, $m_{1}=-0.8$, and $B_{p}=1$~\cite{Matsumoto1985}.  The attractor of this system contains two loops, each of which surrounds an unstable fixed point.  Figure~\ref{fig:DS_Attractor} shows a typical trajectory tracing out the attractor of this system as well as the positions of two control planes.  The control planes are assigned the values $0$ and $1$, and, as the chaotic system traces out a trajectory, a binary sequence can be recorded that we will call the \emph{visitation sequence}.

\begin{figure}[!ht]
	\begin{centering}
		\includegraphics[width=350pt]{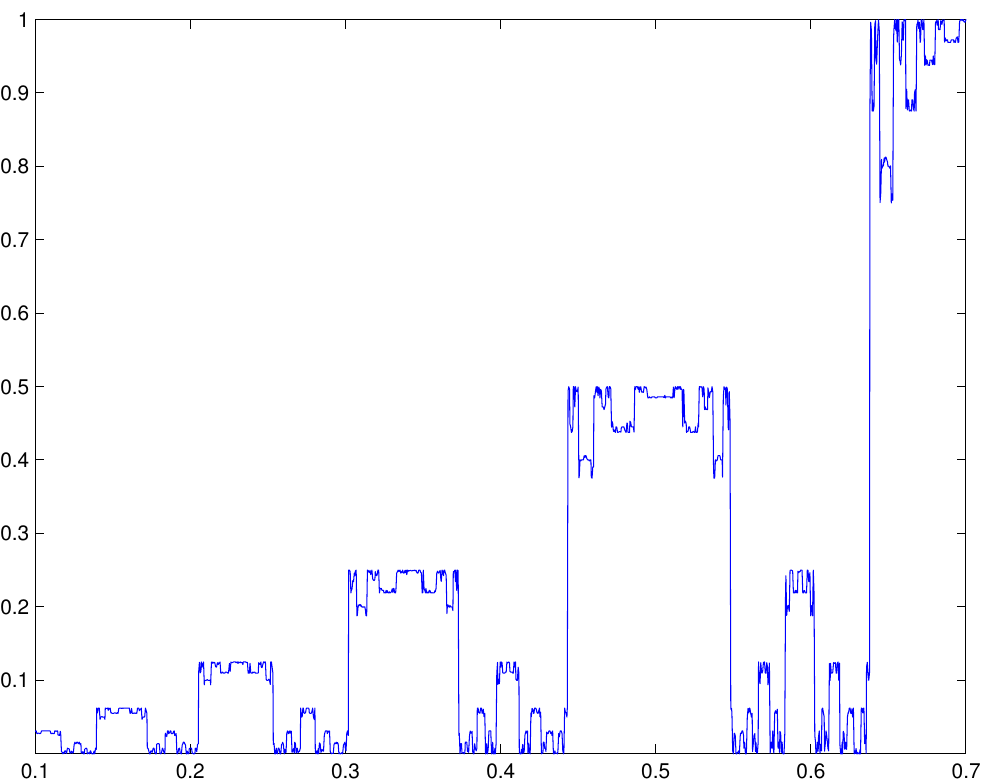}
		\caption{Coding function, $r(x)$.}
		\label{fig:codingfunc}
	\end{centering}
\end{figure}

The first step in producing cupolets is to define a Poincar\'{e} surface of section on each lobe by 
\begin{equation*}
	i_{L} = \pm GF \quad \textrm{and} \quad \left\vert v_{C_{1}}\right\vert \leq F,
\end{equation*}
where $F = \frac{B_{P}\left(m_{0}-m_{1}\right)}{\left(G+m_{0}\right)}$.  The two Poincar\'{e} sections are used as control planes whenever trajectories of this system are being controlled.  One edge of each control plane intersects the attractor at an unstable fixed point at the center of each lobe as seen in Figure~\ref{fig:DS_Attractor}.  Even though the Poincar\'{e} surface is two-dimensional, the attractor is locally nearly two-dimensional.  This means that the intersection between the Poincar\'{e} surface and the attractor is approximately one-dimensional which allows us to partition the intersection into $M$ equally spaced bins.  The distance from the center of each bin to the center of the corresponding lobe is recorded and each bin center is then used as a starting point from which the chaotic system is simulated without control.  After $N$-many loops around the attractor, one obtains a binary lobe visitation sequence for each trajectory.  Such a sequence is represented by a coding function, $r_{N}(x),$ which maps an initial condition, $x$, on either section to a binary representation of its future visitation sequence.  More precisely, if $x$ results in the visitation sequence $b_{1}b_{2}b_{3}\ldots$, this sequence is mapped to the binary decimal $0.b_{1}b_{2}b_{3}\ldots $ with $r_{N}(x)$ defined by
\begin{equation*}
	r_{N}(x) = \sum_{n=1}^{N}\frac{b_{n}}{2^{n}}.
\end{equation*}
Figure~\ref{fig:codingfunc} shows a plot of this function.  Once $r_{N}(x)$ has been calculated for every bin center on both Poincar\'{e} sections, the future visitation sequence of any point on the Poincar\'{e} surface is known for $N$-many iterations.

To control the system to follow a desired visitation sequence, we run the simulation until its trajectory crosses a Poincar\'{e} section, say at $x_{0}$.  The goal is to steer the trajectory through $x_{0}$ so that in the future, the trajectory traces out a desired visitation sequence represented by $r_{N}(x)$.  If the last digit of the visitation sequence of $x_{0}$ is the same as the first digit of the desired sequence, we let the trajectory pass through the control plane (essentially) unperturbed and continue to evolve until its next intersection with the Poincar\'{e} surface.  Otherwise, we can search $r_{N}(x)$ for the nearest bin center having the same visitation sequence as $x_{0}$ except at the last digit.  More precisely, we locate an $x_{1}$ on the same Poincar\'{e} section as $x_{0}$ with $\left| r_{N}(x_{0})-r_{N}(x_{1})\right| = 2^{-N}$ and then perturb the trajectory to start at $x_{1}$.  These perturbations are known as \emph{macrocontrols} and are guaranteed to be small.

In every unperturbed situation, the trajectory is reset to the center of the bin where it intersects the control plane.  This resetting process may be thought of as applying \emph{microcontrols}.  Applying microcontrols not only minimizes the accumulation of round-off error and the system's sensitivity to initial conditions, but also suggests that some cupolets may intersect.  As will be discussed in Section~\ref{subsec:application_of_cupolets}, this intersection would occur exactly in the center of a bin.

\begin{figure}[!ht]
	\begin{centering}
		\includegraphics[width=350pt]{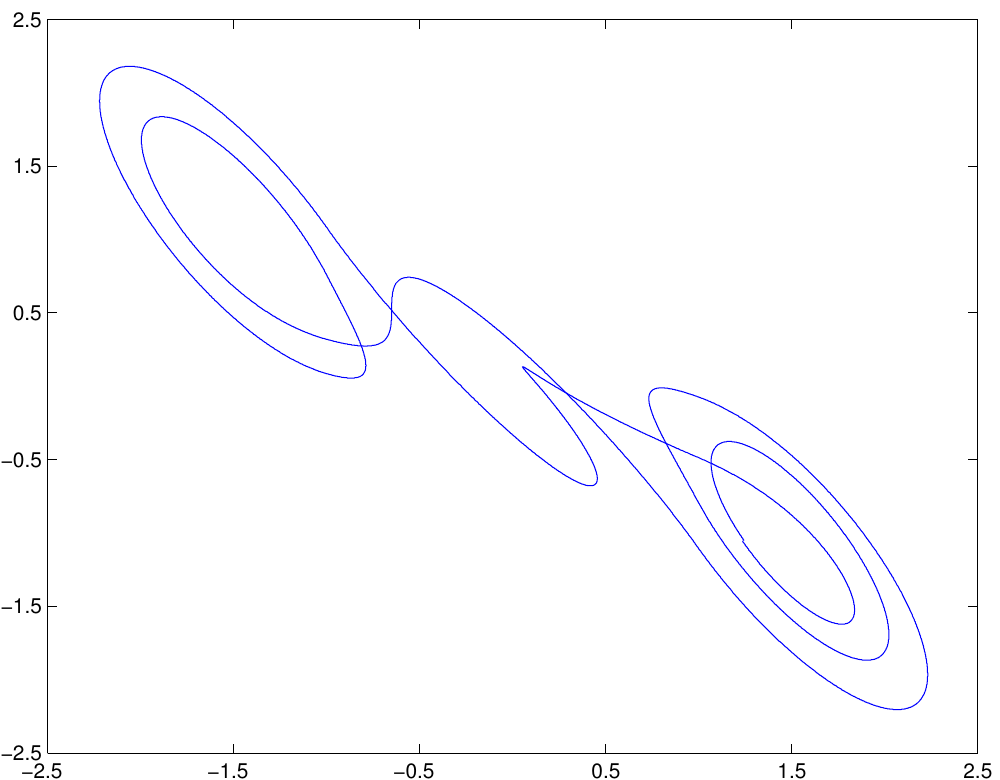}
		\caption{The period 5 cupolet $\mathbf{C}00001$ having the visitation sequence $\mathbf{V}11001$.}
		\label{fig:C00001_period_5}
	\end{centering}
\end{figure}

More generally, the process of controlling the system occurs whenever a trajectory crosses one of the control planes.  The option to perturb the trajectory or allow it to pass freely through the control plane is described by a binary control code with $1$ meaning apply a macrocontrol and $0$ meaning apply only a microcontrol.  In almost all cases, if the controls are applied repeatedly, the system will stabilize onto a unique periodic orbit regardless of its initial state.  These approximate chaotic unstable periodic orbits are known as \emph{cupolets} \cite{Parker1999, Zarringhalam2007}.  Figure~\ref{fig:C00001_period_5} shows a period-5 cupolet.  For organizational purposes, we name the cupolets according to the control codes used to generate them.  The cupolet shown below is stabilized by repeatedly applying the control code `00001' and is named $\mathbf{C}00001$.  The naming scheme for a cupolet's visitation sequence is similar: the visitation sequence of this cupolet is `11001', or $\mathbf{V}11001$.  Visitation sequences either have the same length as their cupolet's control code, or are an integer multiple of the length of the control code.

We implement this control technique in order to produce stabilized cupolets which are essentially approximate periodic orbits of the chaotic systems and whose dynamical behaviors are easily accessible via small controls \cite{Parker1999, Zarringhalam2007, Johnson2009}.  That these orbits are thus produced from such small perturbations suggests that they are shadowing true periodic orbits and theorems have been developed to establish conditions under which this holds \cite{Zarringhalam2007}.   This technique allows a finite, but sufficiently large number of cupolets to be generated regardless of the initial state of the system.  For example, over $8,000$ cupolets have been created from the double scroll oscillator using $16$ bit control codes.  Several other chaotic systems such as the Lorenz and R\"{o}ssler systems have also been used to generate cupolets, as have general unimodal maps.

\subsection{Applications of Cupolets}
\label{subsec:application_of_cupolets}

\begin{figure}
	\begin{centering}
		\includegraphics[scale=0.5]{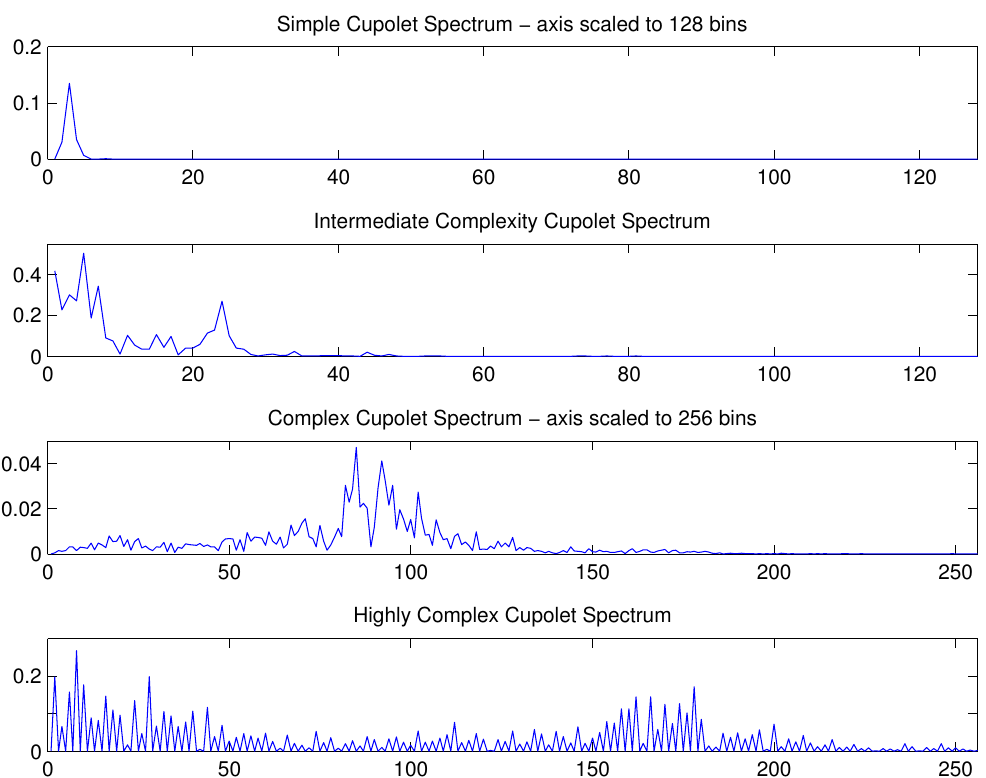} 
		\caption{Spectral variation among cupolets \cite{Short2005a}.}
		\label{fig:cupolet_freqdomain_variation}
	\end{centering}
\end{figure}

\begin{figure}[!ht]
	\begin{centering}
		\includegraphics[scale=0.5]{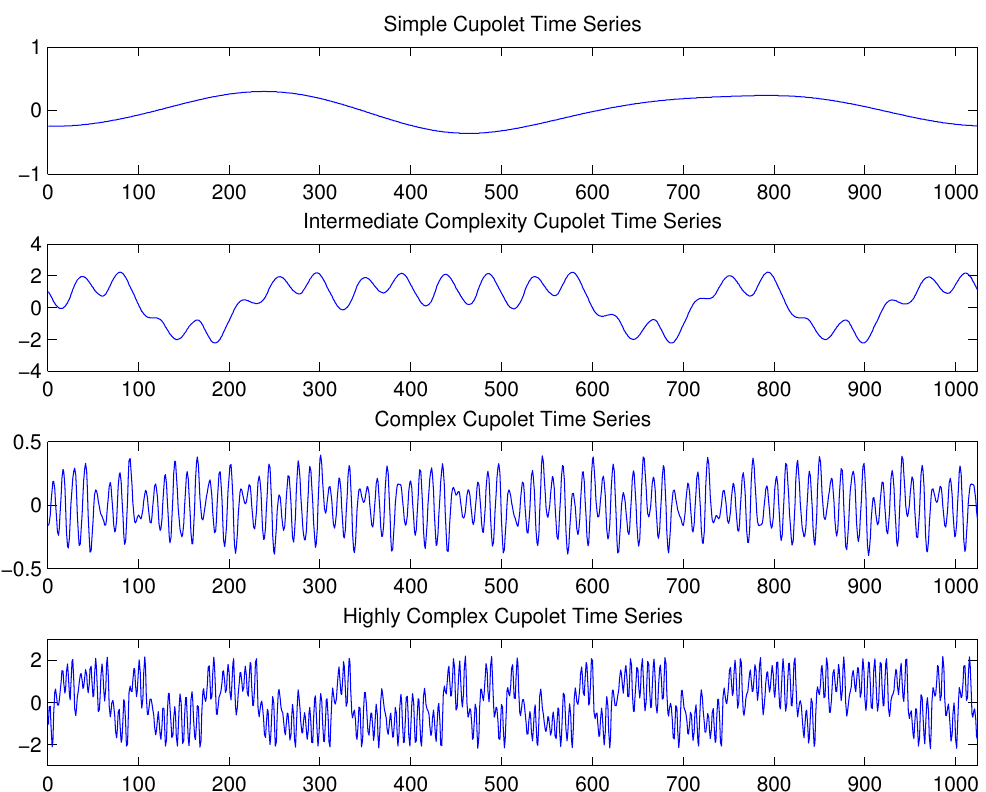} 
		\caption{Time-domain variation among cupolets \cite{Short2005a}.}
		\label{fig:cupolet_timedomain_variation}
	\end{centering}
\end{figure}

Cupolets have been found particularly useful in applications not only in secure communication \cite{Parker1999, Short1999}, but also in image processing and data compression.  At a fundamental level, cupolets are very rich in structure and may be used to generate a variety of different waveforms ranging from a simple \emph{sine}-like wave with a single dominant spectral peak to more involved waveforms consisting of many harmonics.  Figure~\ref{fig:cupolet_freqdomain_variation} illustrates the high diversity in spectral signatures found among cupolets.  The data is taken from the FFT of a single period of oscillation of each cupolet.  The corresponding cupolets in the time domain can be seen in Figure~\ref{fig:cupolet_timedomain_variation}.  It is clear from comparing the figures that the simplest cupolet in part (a) of each figure is essentially sinusoidal while much richer structure is apparent from parts (b-d).  Cupolets may thus be used to produce signals by using cupolet spectra to replace (part of) a signal's spectrum.  Each of these cupolets may be produced using very little information; we typically use 16 bit control codes.  This makes cupolets especially suited for signal approximations and data compression where efficiently representing real-world signals (e.g., audio and video) is important.  Using more complex signals such as cupolets can aid in achieving faster convergence rates. 

For example, in \cite{Short2005a, Short2005b}, Short et al.\ describe a high quality, low-bitrate audio compression technology, one of the first practical applications of cupolets.  Their compression methods rely on the broad spectral properties of cupolets when reconstructing an audio signal after it has been decomposed into its signal components.  Cupolets with the spectral characteristics closest to a given signal component are chosen and adjusted if necessary in frequency, phase, and amplitude for the reconstruction.  This allows cupolets to reproduce an audio sample since each bin in the sample's FFT may be represented by a single cupolet.  The original signal may ultimately be rebuilt at a very low bitrate without significantly compromising signal quality.  Practical consequences of this technology include efficient transfer of music over mobile networks, file storage on portable devices, and digital rights management.

One significant theoretical result is found in an investigation of Zarringhalam and Short who demonstrate how to transform cupolets from their natural oscillatory state resembling the sinusoidal basis of Fourier analysis into a \emph{compact cupolet} state \cite{Zarringhalam2007}.  Compact cupolets are wavelet-like since their energy is concentrated instead around a single maximum.  From there, an adaptive basis for the space of real-valued functions of a discrete variable may be constructed from only a few compact cupolets.  Fourier and wavelet transforms are two common techniques used in image compression, but compact cupolets combine the most useful features of both methods.

\begin{figure}[!ht]
	\begin{tabular}{cc}
		\resizebox{0.45\textwidth}{!}{\includegraphics{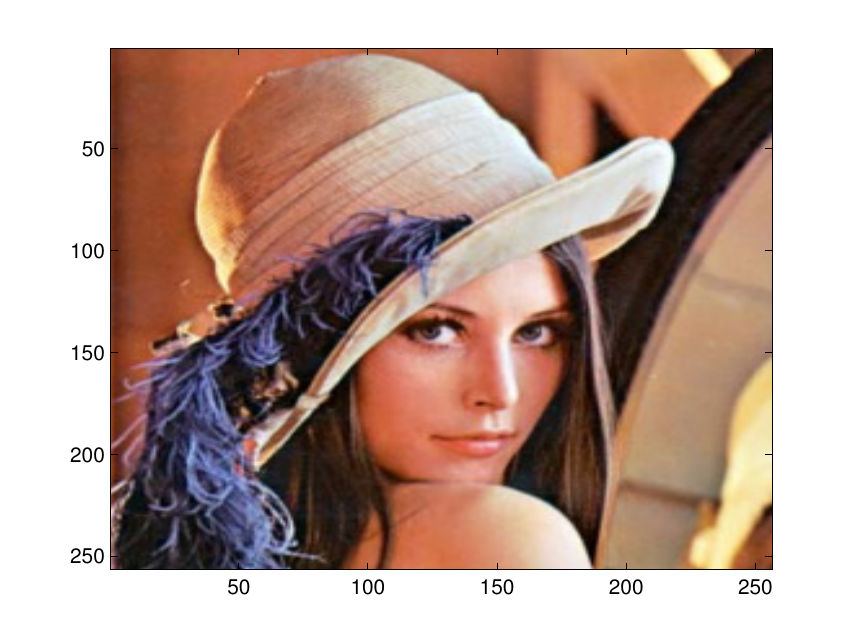}} &
		\resizebox{0.45\textwidth}{!}{\includegraphics{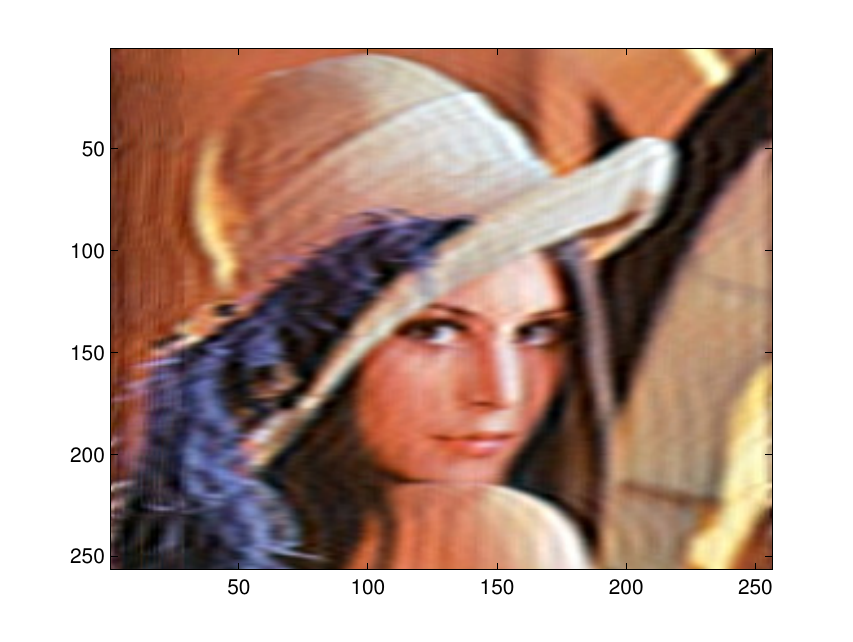}} \\
		(a) & (b) \\
		\resizebox{0.45\textwidth}{!}{\includegraphics{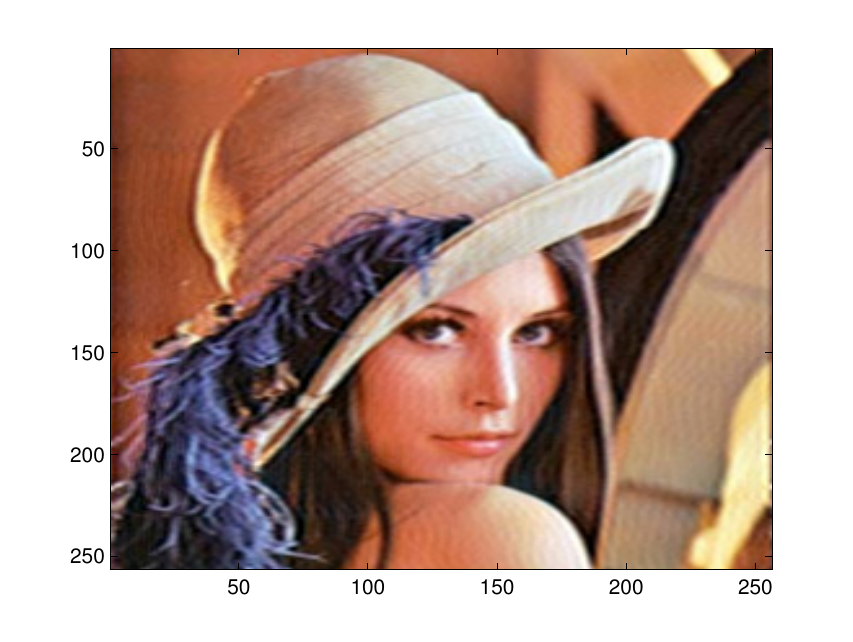}} &
		\resizebox{0.45\textwidth}{!}{\includegraphics{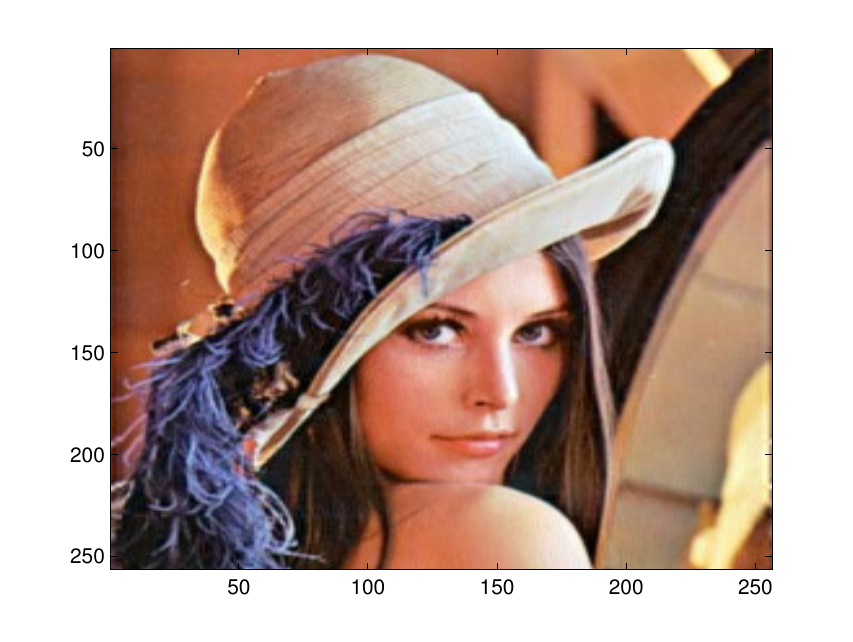}} \\ 
		(c) & (d) \\
	\end{tabular}
	\caption{(a) Original $256\times 256$ image and incremental resolution levels.  The number of basis elements in each
	resolution level is (b) $24$ (c) $56$ and (d) $120$ per window \cite{Zarringhalam2008}.}
	\label{fig:PPMRA_lenna}
\end{figure}

Zarringhalam and Short then extended this multiresolution analysis using compact cupolets to image compression \cite{Zarringhalam2008}.  The compressions of a sample image using compact cupolets are shown in Figure~\ref{fig:PPMRA_lenna}.  The size of the image is $256\times 256$ pixels and the size of the data window is a single line scan, or $256$ points.  The number of basis elements per data window is $24,56,$ and $120$ respectively.  At the third resolution level with $56$ basis elements, the reconstructed image is essentially perfect.  The convergence is extremely rapid when considering such a low number of basis elements are needed.

Additional cupolet research by Johnson and Short demonstrates how to transition between cupolets and ultimately move efficiently about a chaotic attractor \cite{Johnson2009}.  This result takes advantage of two important properties of cupolets.  The first property is that cupolets are generated independent of initial conditions, making it possible to cause the system to stabilize onto a desired cupolet from any initial point.  Second, when a chaotic trajectory is being directed onto a cupolet by the previously described control scheme, macrocontrols or microcontrols are applied whenever the trajectory passes through a Poincar\'{e} section.  This resets the trajectory to the center of a bin, so some cupolets may intersect at exactly these bin centers.  If two cupolets do intersect, Johnson establishes that switching between them is a smooth and continuous transition.  If they do not intersect, the transition may involve significant transients if it is not carefully controlled.

\begin{figure}[!ht]
	\begin{centering}
		\begin{tabular}{cc}
			\resizebox{0.485\textwidth}{!}{\includegraphics{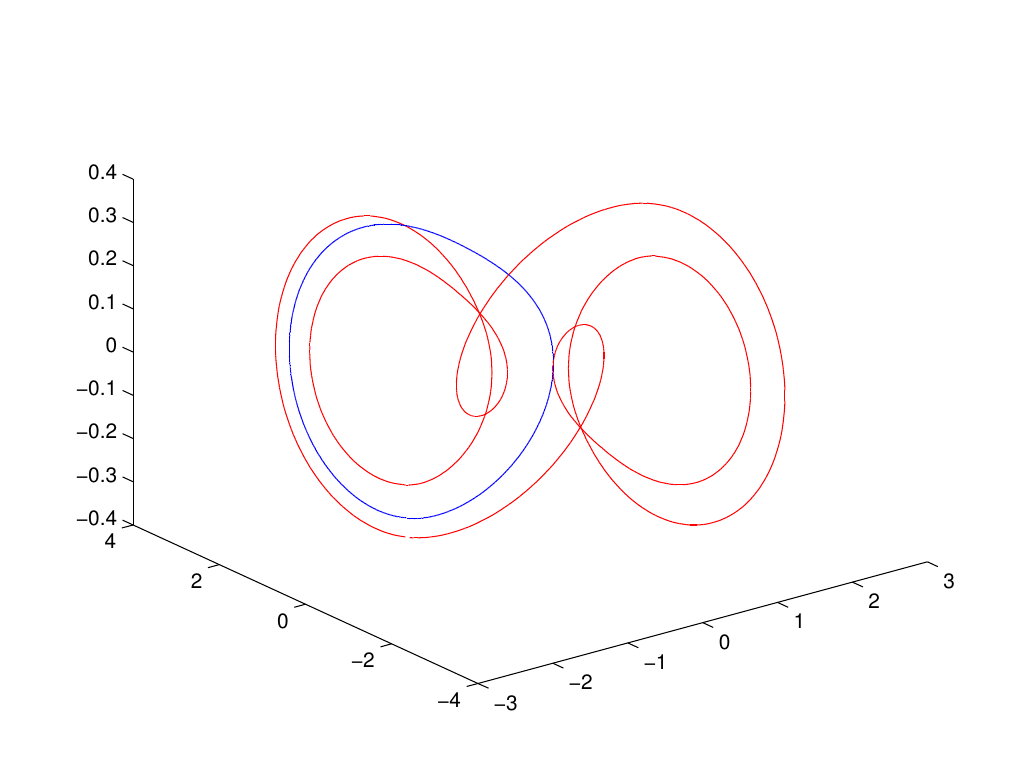}} 
			&
			\resizebox{0.485\textwidth}{!}{\includegraphics{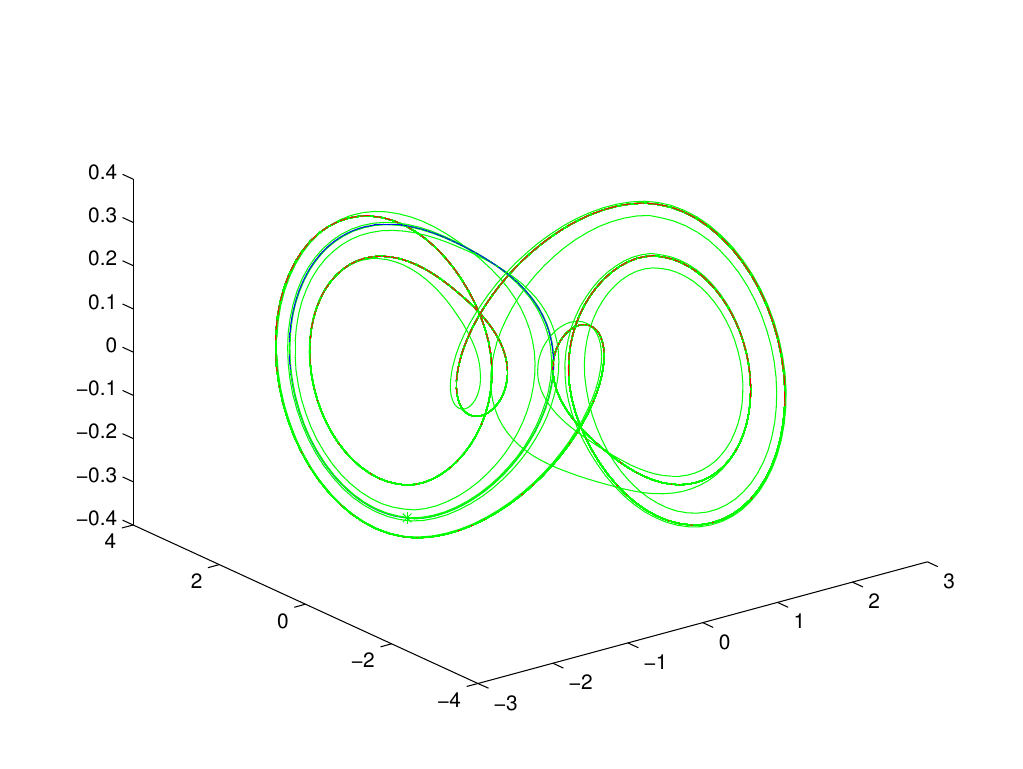}} \\
			(a) & (b) \\
		\end{tabular}
		\caption{Cupolet transition with (a) zero transient and (b) non-zero transient.  The blue and red orbits are
		cupolets while the green curve in (b) is the chaotic transient \cite{Johnson2009}.}
		\label{fig:intersecting_cupolets}
	\end{centering}
\end{figure}

The procedure for transitioning between two intersecting cupolets, say cupolets A and B, is to ride cupolet A until its intersection with cupolet B at the center of a particular bin on a Poincar\'{e} section.  To switch to cupolet B, the control code for cupolet B must be applied using the control scheme at the bin where the intersection occurs, whereupon the chaotic transition leaves the first cupolet in a continuous manner.  The control sequence for cupolet B is then periodically repeated and the switch to cupolet B is complete.  The results of switching between two intersecting cupolets can be seen in Figure~\ref{fig:intersecting_cupolets}(a).  If the control codes for cupolet B are applied in the wrong (non-intersecting) bin, then a long transient occurs before the system stabilizes onto cupolet B, as illustrated in Figure~\ref{fig:intersecting_cupolets}(b).

\vspace{-0.25cm}

If cupolets A and B do not intersect at any bin, the abundance of cupolets suggests that it may be possible to obtain an intermediary sequence of intersecting cupolets such that cupolet B may be realized from cupolet A by transitioning between each cupolet in the sequence.  Algebraic graph theory can be used to determine the optimal series of intersecting cupolets, C,D,E, $\ldots\,,$ and corresponding intersection bins, $\alpha, \beta, \gamma, \ldots\,,$ that should be used in order to minimize the transition transients.  A directed digraph may be created where vertices represent bins and an edge represents the cupolet connecting the two bins (vertices) while passing through the fewest number of intermediate bins.  The number of bins a cupolet must pass through in order to visit the bins (vertices) it joins is given as the weight of each edge.  Dijkstra's shortest path algorithm may then be used to determine the ideal cupolet transition sequence that reveals the shortest sequence of intermediate cupolets needed to switch from cupolet A to cupolet B.  Thus, to transition from cupolet A to cupolet B, the path would jump from A to C at the intersection bin $\alpha$, ride cupolet C until intersecting with D at $\beta,$ then E at $\gamma$ and so forth until finally switching onto B.



\section{Entangled Cupolets}
\label{sec:entangled_cupolets}

In this section, we discuss a way to entangle two chaotic systems by controlling a pair of their cupolets into a state of mutual stabilization.  Since a given control code uniquely identifies a cupolet, cupolets which we consider entangled have dynamics that interact with an exchange function so that their oscillations become mutually self-supporting.  Any slight disturbance in the interaction may be enough to compromise their entanglement.  This evokes a reasonable analog to that of quantum entanglement which will be further explored in a future paper~\cite{Morena2012}.

\begin{figure}[!ht]
	\begin{centering}
		\begin{tabular}{cc}
			\includegraphics[scale=0.42]
			{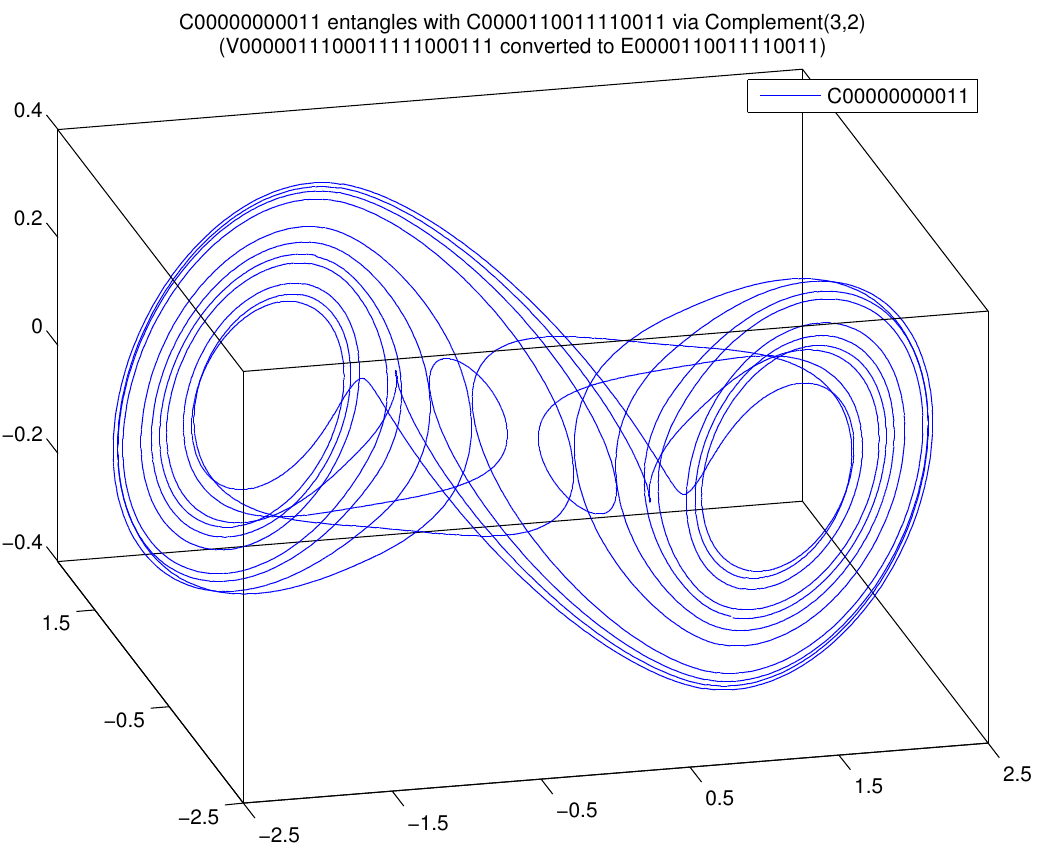} &
			\includegraphics[scale=0.42]
			{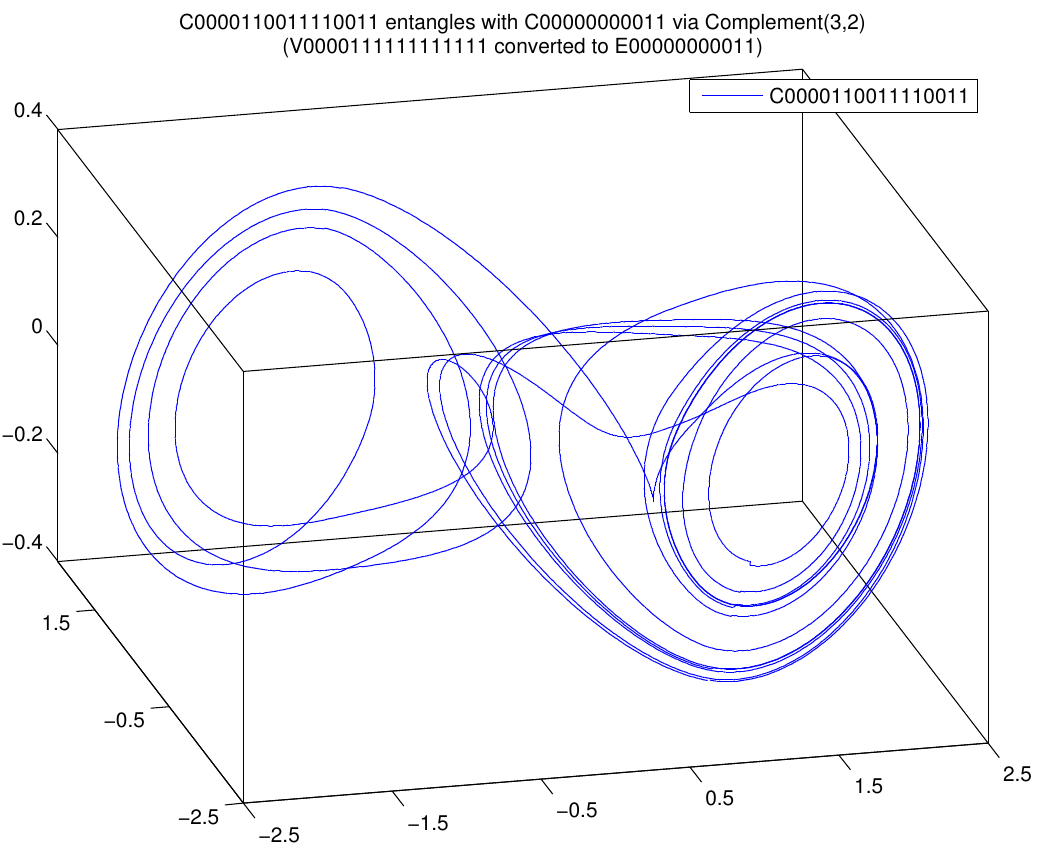} \\
			(a) & (b)
		\end{tabular}
		\caption{Entangled cupolets (a) $\mathbf{C}00000000011$ and (b) $\mathbf{C}0000110011110011$ with visitation sequences 	
		$\mathbf{V}0000011100011111000111$ and $\mathbf{V}0000111111111111$, respectively.}
		\label{fig:entanglement_example_C00000000011_C0000110011110011}
	\end{centering}
\end{figure}

Our method of chaotic entanglement is perhaps best illustrated with an example.  To generate entanglement between the cupolets shown in Figure~\ref{fig:entanglement_example_C00000000011_C0000110011110011}, two double scroll oscillators, Systems I and II, are first simulated without control.  In order to stabilize one of the cupolets, say $\mathbf{C}00000000011$, the control code `00000000011' must be applied to System I.  After $\mathbf{C}00000000011$ completes one full period around the attractor, we obtain its visitation sequence,  $\mathbf{V}0000011100011111000111$.  The visitation sequence is then passed to an exchange function where it is modified according to a predefined binary operation and sent to the second chaotic system as an \emph{emitted sequence}.  In this example, a `complement' exchange function converts $\mathbf{V}0000011100011111000111$ into the emitted sequence $\mathbf{E}0000110011110011$ by interchanging subsequences of ones and zeros for zeros and ones.  Exchange functions will be described more fully in our forthcoming paper~\cite{Morena2012}.  Once it is passed out of the exchange function, an emitted sequence is applied to the second chaotic system, but as a control code that possibly stabilizes System II onto another cupolet.  In this case, system II stabilizes onto the other cupolet, $\mathbf{C}0000110011110011$, since the emitted sequence, $\mathbf{E}0000110011110011$, \emph{actually} is this cupolet's control code.

At this point, the cupolet interaction is repeated in the reverse direction.  After evolving one full period about the attractor, the visitation sequence, $\mathbf{V}0000111111111111$, of cupolet $\mathbf{C}0000110011110011$, is recorded and passed to the same exchange function.  The sequence is emitted as $\mathbf{E}00000000011$ and is then applied to System I as a control code.  Because $\mathbf{E}00000000011$ matches the control code, `00000000011', needed to generate the first cupolet, $\mathbf{C}00000000011$, this cupolet's stability is preserved.  In fact, both emitted sequences, $\mathbf{E}00000000011$ and $\mathbf{E}0000110011110011$, match the required control codes for cupolets $\mathbf{C}00000000011$ and $\mathbf{C}0000110011110011$, respectively.  This defines the cupolet's interaction which has been managed by the exchange function.  If the interaction were to continue, then each cupolet's stability would be guaranteed and we thus consider the pair to be \emph{entangled}.  This particular entangled cupolet example is illustrated in Figure~\ref{fig:entanglement_example_C00000000011_C0000110011110011}.

On a less technical level, a pair of cupolets, A and B, chaotically entangles once each cupolet's visitation sequence mutually preserves the other's stabilization.  As cupolet A winds its way about the lobes of its attractor, the bits of its visitation sequence are passed to the exchange function which produces the emitted bits that then can go on to control cupolet B.  In the reverse direction, with the help of the exchange function, cupolet A receives the necessary control code bits needed to maintain its stabilization (since the visitation sequence belonging to cupolet B passes through the exchange function and is emitted as control bits for cupolet A).  Subtle disruptions in the message exchange will almost always destroy the entanglement, unless a great deal is known about the control scheme.  For instance, consider the process of making a measurement on an entangled system that has the effect of interchanging a 0 for a 1, or vice versa.  A cupolet's control code is unique so if such a disturbance is made to cupolet A's control code, at any stage of the entanglement process, then either cupolet A destabilizes or it transitions to a different cupolet as is described in Section~\ref{subsec:application_of_cupolets}.  The effect in both cases is to produce a different visitation sequence that no longer guarantees the stability of cupolet B, and so their entangled state is lost.

The entangled pair's interaction is managed by the exchange function whose role essentially is to modify visitation sequences and emit them as cupolet-stabilizing control codes.  The entanglement is self-perpetuating and no further controls need be applied.  We are continuing to experiment with various types of exchange functions, motivated by implementation constraints or potential applications.  For example, the use of an `integrate-and-fire' exchange function is inspired by various investigations of chaos in neuron models \cite{Racicot1997, Erichsen-Jr2006}.  In this scenario, the bits of a visitation sequence are interpreted as binary energy which are accumulated in the bins of a bit register.   Each bin may represent a time increment of the neural network.  As the register slides along the visitation sequence, bits of the visitation sequence are placed one at time into their corresponding register bin and their numerical sum is calculated.  When the cumulative energy inside the register `integrates' to a threshold level, the system `fires' and the exchange function emits a 1.  For each time interval the register energy remains below the threshold, the exchange function emits a 0 and the register advances one bit in the visitation sequence.  After a `fire', we reset the register bins to zero.

Another type of exchange function which has generated cupolet entanglement is a `preponderance' type function, whereby bits in the visitation sequence are again taken as binary energy.  If the cumulative energy of a portion of a visitation sequence reaches a threshold level, that entire subsequence is replaced by a bit string composed entirely of ones.  If the threshold is not reached, the portion is instead reset to a bit string consisting only of zeros.  In some sense, this exchange function resembles a discrete finite impulse response (FIR) filter from signal processing, since an output (cumulative energy) is obtained from a weighted sum of the current and previous bits (energies) of a visitation sequences.

\begin{figure}[!ht]
	\begin{centering}
		\begin{tabular}{cc}
			\includegraphics[scale=0.42]
			{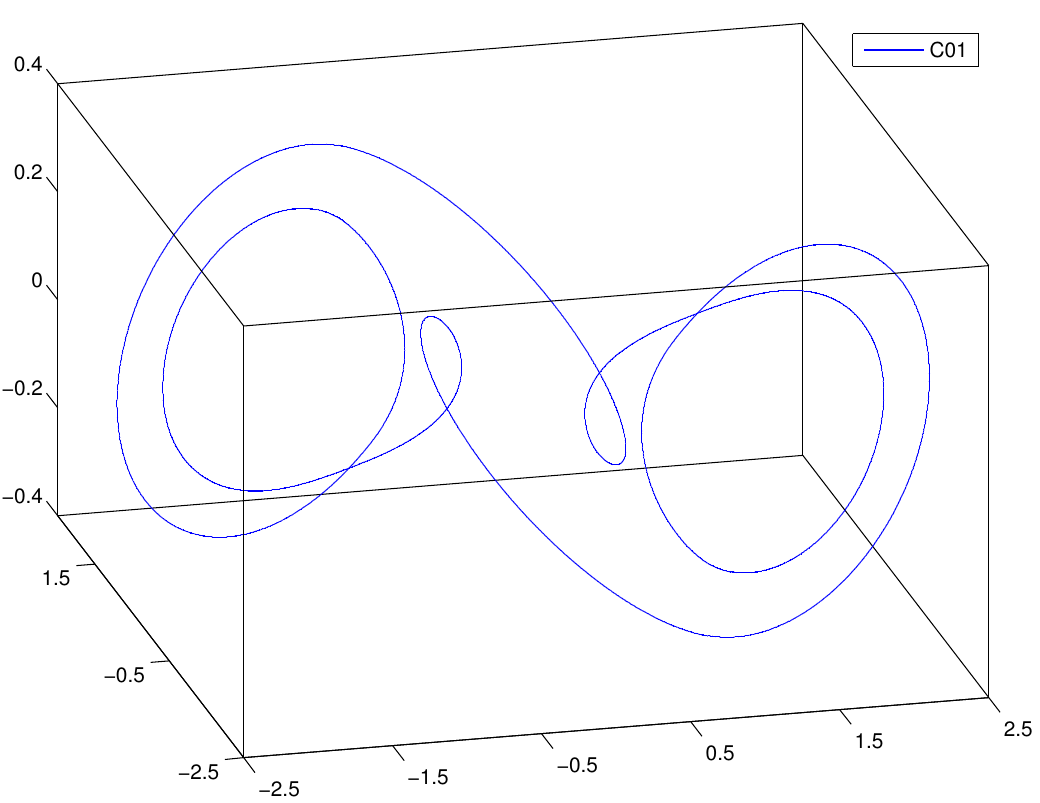} & 
			\includegraphics[scale=0.42]
			{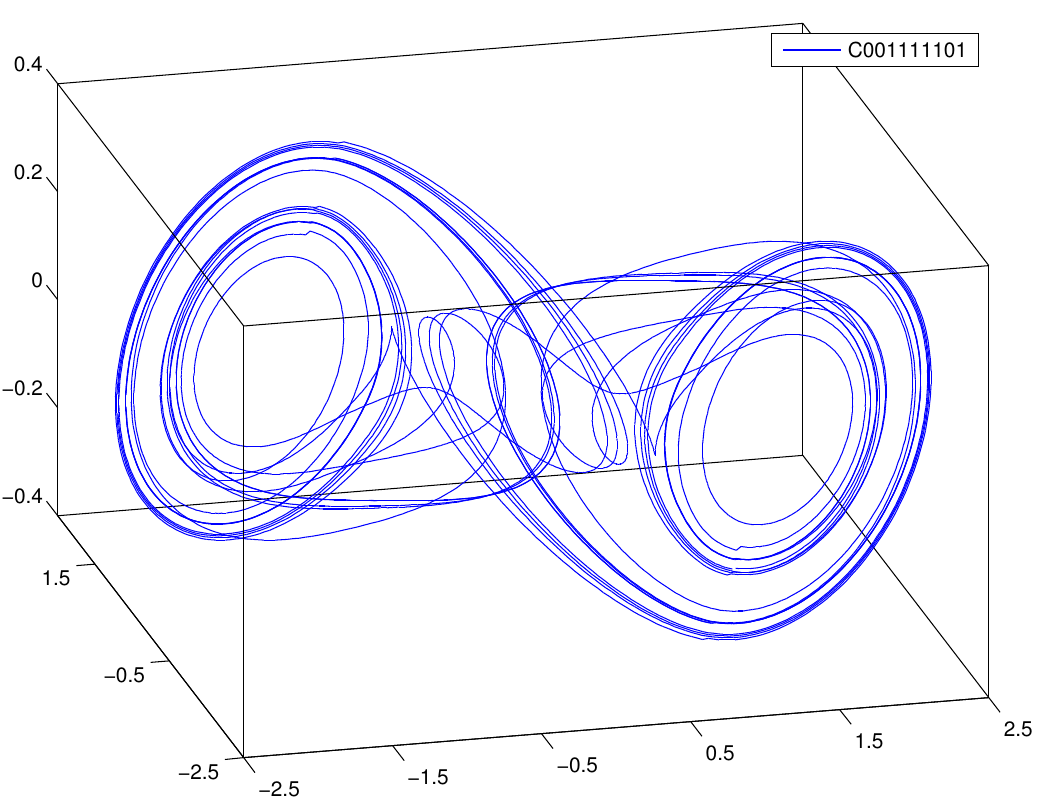} \\
			(a) & (b) \\
			\includegraphics[scale=0.42]
			{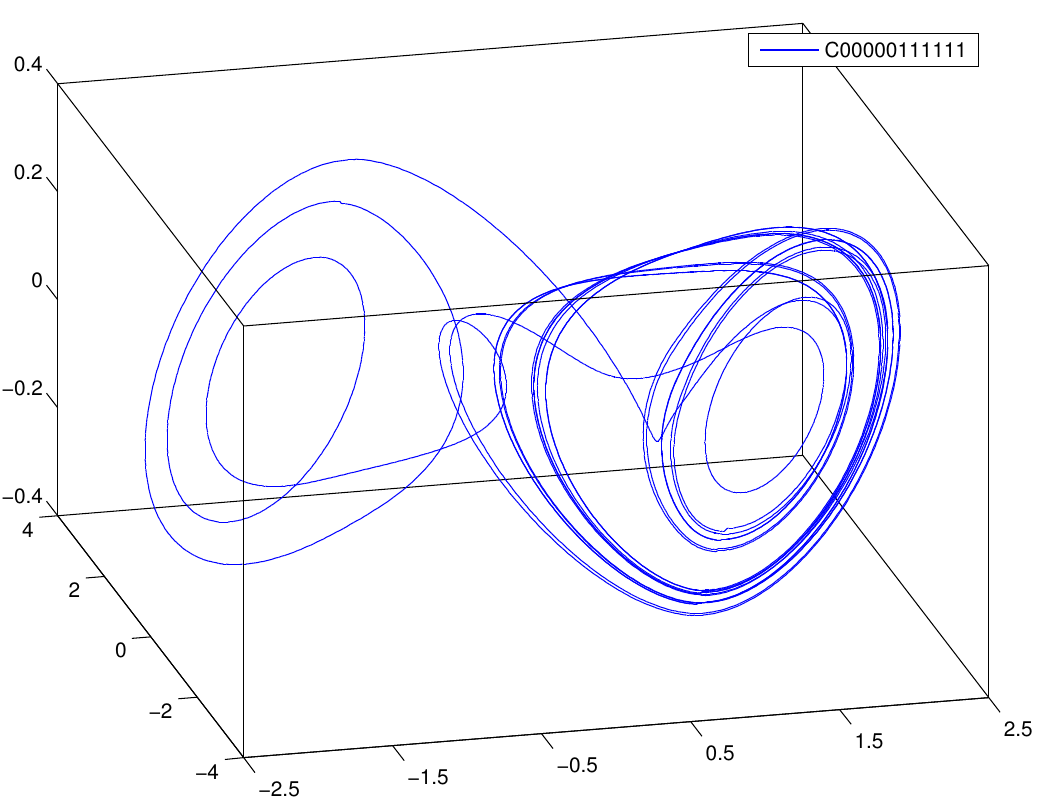} &
			\includegraphics[scale=0.42]
			{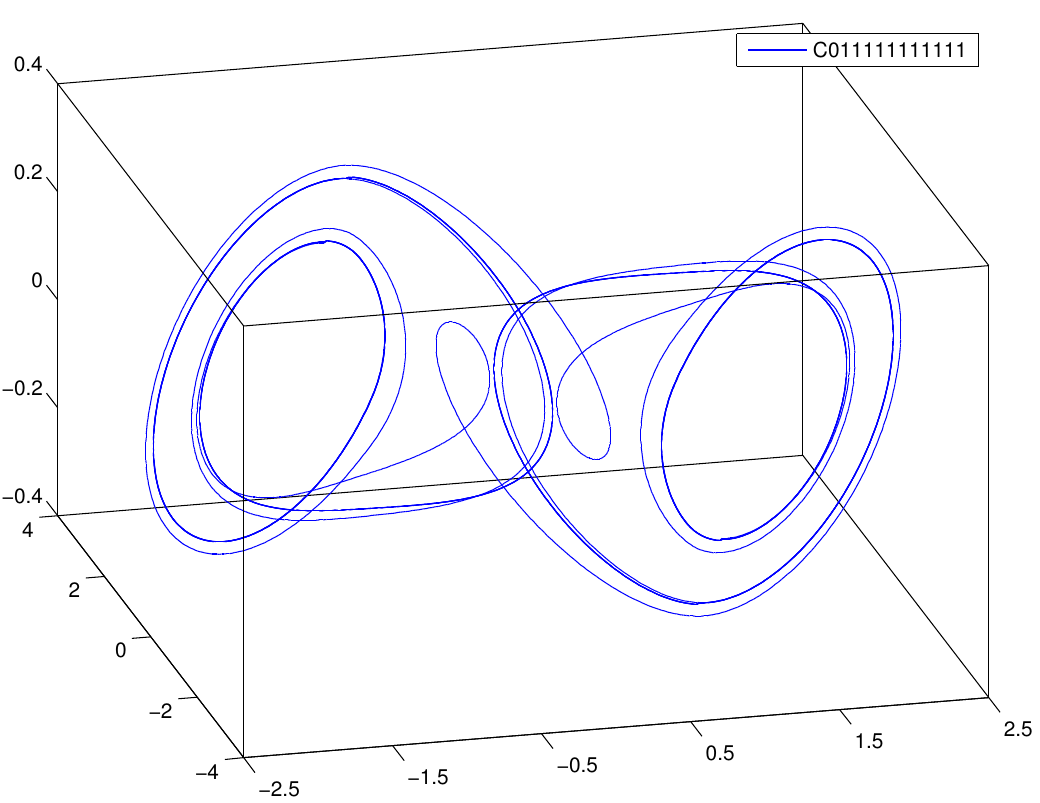} \\
			(c) & (d) \\
		\end{tabular}
		\caption{Various entangled cupolets.  The original visitation sequences are (a)~$\mathbf{V}0011$, 
		(b)~$\mathbf{V}0000000111110011111110000011$, (c)~$\mathbf{V}0001111111111111111111$, and 
		(d)~$\mathbf{V}000000000000111111111111$.}
		\label{fig:entangled_cupolets}
	\end{centering}
\end{figure}

More technical details of our entanglement procedure are included in a forthcoming paper \cite{Morena2012}.  At this time, we have investigated the double scroll system for entanglement by using over 350 variations of exchange functions in 8 major categories.  We have successfully observed entanglement between over 150 cupolets and have classified certain types of entanglement.  \emph{Cross-entanglement} refers to entanglement between two different cupolets while entanglement between the same cupolet is called \emph{self-entanglement}.  Figure~\ref{fig:entangled_cupolets} shows a few entangled cupolets along with their control sequences.  The cupolets in (a) and (b) exhibit self-entanglement, while cupolets (c) and (d) cross-entangle with each other.  Entangled cupolets need not share the same periods, as is demonstrated by the entangled cupolets of Figure~\ref{fig:entangled_cupolets}(c, d), which are of periods 22 and 24, respectively.

\section{Future Research} 
\label{sec:future_research}

In future research, we anticipate several additional areas of investigation.  First, since it is possible to transition from one cupolet to another through an intermediate series of intersecting cupolets, it may be possible to identify a subset of key cupolets that, in some sense, span the space of dynamical behaviors on the attractor.  We will then investigate whether it is possible to decompose a higher-order cupolet into a combination or `weighted sum' of simpler cupolets.  We speculate that from any large set of cupolets, one can numerically find a subset of `fundamental' cupolets, analogous to a basis set of vectors.  An \emph{expansion with respect to cupolets} could therefore be defined.

Second, our present notion of entanglement is space-independent, meaning that two chaotic systems may entangle in spite of the separation distance between them.  As an alternative, we plan to investigate cupolet entanglement when the controls and the cupolet interaction are functions of the separation distance.  Other control techniques that do not involve the $r_{N}(x)$ coding function will also be considered.  If similar chaotic entanglement can be found, it may be more easily produced in physically realizable systems.

Another potential area for investigation would be to extend these techniques to the `kicked top' model that was mentioned earlier in Section~\ref{sec:introduction}.



\section{Conclusions}
\label{sec:conclusions}

We hope that the rich structure of cupolets will lead to many additional physical applications.  We have discussed how effective cupolets can be at representing complex signals, making them particularly useful in image processing and data compression.  We have also seen how transitioning between cupolets makes it possible to move efficiently around a chaotic attractor. The potential thus exists for cupolets to improve the interchange between chaotic systems and other research areas, but there is much more work left to be done.

In particular, we have presented an analog of (quantum) entanglement in cupolets.  This entanglement occurs independent of the distance between the chaotic systems, but is sensitive to measurement disturbances.  Initial results are promising, since several hundred entangled cupolets have been found.  Future research will investigate alternative entanglement procedures as well as search within a large set of cupolets for a fundamental subset.

In summary, the study of periodic orbits of chaotic systems continues with cupolets and appears very promising from both a theoretical and a practical perspective.  It will be interesting to see if producing periodicity from chaos can play a key role in additional developments.



\bibliography{morena-short-entangled-cupolets-paper-2012_references}
\bibliographystyle{unsrt} 

\end{document}